\documentclass[prb,aps,twocolumn,groupedaddress,floats,showpacs,final,superscriptaddress]{revtex4-2}
\usepackage[latin3]{inputenc}
\usepackage[makeroom]{cancel}
\usepackage{graphicx}
\usepackage{amsmath}
\usepackage{amsfonts}
\usepackage{amssymb}
\usepackage{chngcntr}
\usepackage{color}
\usepackage[left=2cm,right=2cm,top=2cm,bottom=2cm]{geometry}

\usepackage{graphicx} 
\usepackage{dcolumn}
\usepackage{bm}
\usepackage{simplewick}
\usepackage{array}
\usepackage{appendix}

\RequirePackage[
   hyperindex,colorlinks,bookmarksnumbered,
   plainpages=true,pdfstartview=FitH]{hyperref}
\hypersetup{linkcolor=blue,urlcolor=blue,citecolor=blue}
\usepackage{hyperref}
\usepackage{mathrsfs}
\definecolor{purple}{rgb}{0.5,0,0.6}

\usepackage{ulem}
\renewcommand{\emph}[1]{\textit{#1}}
\definecolor{darkblue}{rgb}{0,0,0.5}
\definecolor{darkgreen}{rgb}{0,0.5,0}
\definecolor{darkred}{rgb}{.7,0,0}
\definecolor{purple}{rgb}{0.5,0,0.6}
\definecolor{orange}{rgb}{1,0.5,0}
\definecolor{grey}{rgb}{.6,.6,.6}
\definecolor{lightpink}{rgb}{1,0.7,0.75}
\definecolor{pink}{rgb}{1,0.4,0.58}
\definecolor{deeppink}{rgb}{1,0.08,0.58}

\usepackage{ulem}

\begin{document}

\date{\today}
\title{Heat Coulomb blockade in a double-island metal-semiconductor device}


\author{A. V. Parafilo}
\email{parafilo.sand@gmail.com}
\affiliation{Center for Theoretical Physics of Complex Systems, Institute for Basic Science, Expo-ro 55, Yuseong-gu, Daejeon 34126, Republic of Korea}

\affiliation{Department of Condensed Matter Physics, Faculty of Mathematics and Physics,Charles University, Ke Karlovu 5, CZ-121 16 Prague, Czech Republic}

\date{\today}

\begin{abstract}
We study the thermal transport properties of a mesoscopic device comprising two metallic islands embedded in a two-dimensional electron gas in the integer quantum Hall regime. It is shown that the $2M$ ballistic edge channels connecting the islands to the external reservoirs and the $N$ inter-island channels play a central role in the phenomenon of heat Coulomb blockade. Unlike the single-island case, where the heat flux is reduced by exactly one quantum of thermal conductance, we predict an additional suppression proportional to the factor $M^2/(2N+M)^2$. We further examine a configuration in which the islands are placed between electrodes at different temperatures and identify the conditions under which the Wiedemann-Franz law is violated.

\end{abstract}

\maketitle


Floating Ohmic contacts are small metallic islands strongly coupled to low-dimensional quantum conductors but isolated from external circuitry, exchanging charge and energy only through quantum channels. Typically embedded in highly doped two-dimensional electron gases in the quantum Hall regime, such structures form the core of hybrid metal-semiconductor devices. Acting as local reservoirs, the islands equilibrate charge and energy, regulate decoherence, and enable precise measurements of heat transport~\cite{pierre,Heat_CB,heat,heat2,heat3}. Beyond energy flow, floating contacts serve as versatile probes of quantum phenomena, including dynamical Coulomb blockade~\cite{pierre_DCB,DGG2}, electron state teleportation~\cite{teleport}, and Luttinger-liquid physics~\cite{pierre_LL}. Recent advances have extended this platform to realize multi-channel~\cite{pierre_2CCK,pierre_3CCK,pierre_screening} and multi-impurity~\cite{DGG} Kondo models based on charge pseudospin of metallic island(s)~\cite{matveev,flensberg,furusaki_matveev}, providing a controllable route to explore strongly correlated and collective effects in mesoscopic circuits.


A recent milestone in the study of quantum thermal transport in hybrid metal-semiconductor circuits has been the observation of {\it heat Coulomb blockade}~\cite{Heat_CB}.
In this effect, the heat flux emitted by a metallic island through $N$ ballistic channels shows a universal suppression by one quantum of thermal conductance, $\kappa_0 T$ ($\kappa_0=\pi k_B^2/6\hbar$), which is a direct manifestation of the floating Ohmic contact's finite charging energy $E_{\rm C}=e^2/2C$ ($C$ is the island's capacitance). The importance of this energy scale was clarified in the seminal paper of Slobodeniuk {\it et al.}~\cite{Slobodeniuk}, who formulated a theory of floating Ohmic contacts coupled to quantum Hall edge states. At low temperatures $T\ll E_{\rm C}/k_B$, the charging energy suppresses charge fluctuations on the island, blocking the collective charge mode from contributing to heat transport, so that only $N-1$ neutral modes carry heat, resulting in a total thermal conductance of 
$(N-1)\kappa_0 T$. Equivalently, one can say that the charge mode fails to equilibrate with the island~\cite{Slobodeniuk} and effectively teleports~\cite{Heat_CB,teleport2,teleport} across the contact, preserving its incoming energy.
As temperature increases, charge fluctuations are restored and the blockade is lifted, recovering the full thermal conductance $N\kappa_0T$. Beyond its experimental significance, the discovery of the heat Coulomb blockade has also stimulated theoretical efforts to understand the influence of Coulomb interactions on heat transport phenomena in hybrid mesoscopic systems ~\cite{stabler1,stabler2,stabler3,roche,sela}.


In this Letter, we focus on the thermal properties of the two-site charge Kondo (2SCK) circuit -- an extension of the single-island hybrid metal-semiconductor device to a double-island configuration. Originally introduced in Ref.~\cite{kiselev2018}, this model was proposed as a versatile platform to investigate the competition between Fermi-liquid and non-Fermi-liquid (NFL) regimes through their thermoelectric signatures, as well as to probe the emergence of $\mathbb{Z}_N$ parafermion excitations. Subsequently, the 2SCK circuit was realized experimentally~\cite{DGG} and employed to identify frustrated interactions at an exotic quantum critical point. In particular, the observation of a $\mathbb{Z}_3$ parafermion characterized by the fractional residual entropy $(k_B/2)\log 3$ was reported~\cite{DGG,PhysRevLett.130.146201}.

Following its realization, the 2SCK circuit attracted extensive theoretical attention. Later studies examined its charge~\cite{two-site,PhysRevLett.130.146201, parafilo_2023,PhysRevB.110.L241406,parafilo_fnt}, heat~\cite{nguyen2022,drag,kiselev_WF}, and thermoelectric transport properties~\cite{kiselev2018,PhysRevB.109.115139}, revealing a rich interplay between Coulomb correlations and NFL behaviors. The same architecture has also been recognized as a promising platform for 
simulating strongly correlated one-dimensional phenomena, including tunnel-coupled Luttinger-liquids with different interaction strengths~\cite{parafilo_2023}, resonant tunneling in Luttinger liquid~\cite{PhysRevB.110.L241406}, and the two-channel charge Kondo effect with one Luttinger-liquid channel~\cite{parafilo_fnt,parafilo_ee}. Moreover, theoretical study predicts that the device exhibits a distinctive "magic" Lorenz ratio attributed to the Anderson orthogonality catastrophe~\cite{kiselev_WF}.

Motivated by these predictions, we consider double-island circuits where Coulomb blockade produces two gapped charge modes, independent of the total number of ballistic channels. The magnitudes of these gaps, however, depend sensitively on the channel configuration~\cite{parafilo_2023,parafilo_fnt}, providing experimental means to tune charge and heat transport. This observation leads to several key questions explored in this work: (i) how the heat Coulomb blockade manifests itself in a system of two coupled floating Ohmic contacts; (ii) whether the presence of two gapped modes can give rise to a two-stage blockade; and (iii) how the imbalance between inter-island channels and those connecting each island to external reservoirs affects thermal transport.



\begin{figure}
\centering 
\includegraphics[width=1\columnwidth]{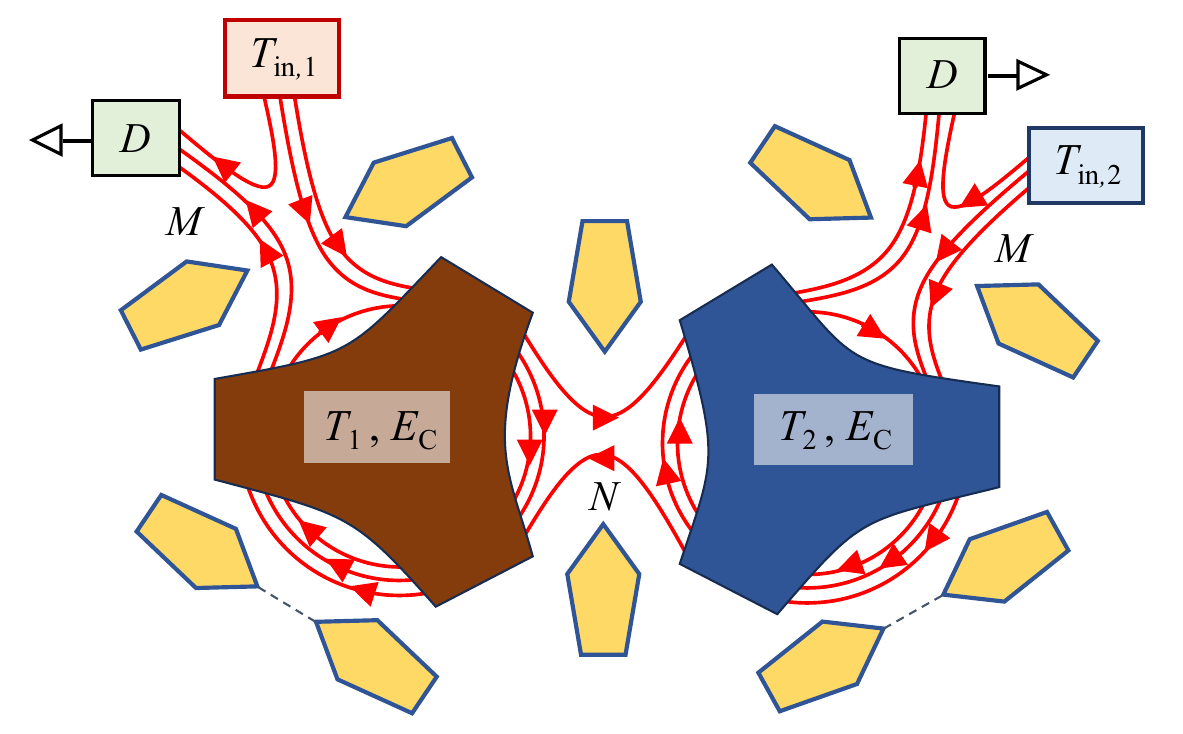} \caption {Schematic representation of the double-island hybrid metal-semiconductor device: two large metallic islands (i.e., two floating Ohmic contacts) are embedded in a two-dimensional electron gas that is set in the integer quantum Hall regime with filling factor $\nu$. Two islands are electrically connected to (i) each other via $N$ ballistic channels, and to (ii) all reservoirs via $2M$ channels. As for example, we show the case of $\nu=3$, $M=2$, and $N=1$. For the heat Coulomb blockade experiment configuration, we assume all reservoirs are at the same base temperatures $T_{\rm in,1}$$=$$T_{\rm in,2}$$=$$T_{\rm in}$. At the same time, two islands are heated up (e.g., by applying dc voltages that dissipate Joule heat into both islands) to the temperature $T_{1}$$=$$T_2$$=$$T_{\rm C}$$>$$T_{\rm in}$. For the heat transport experiment configuration where both islands are placed between heat source and drain, we assume $T_{\rm in,1}$$=$$T_S$, $T_{\rm in,2}$$=$$T_D$, while the temperatures of both islands $T_{1,2}$ are not fixed.}
\label{Fig1} 
\end{figure}

\textit{Model.} In this Letter, we consider a double-island hybrid metal-semiconductor device, first realized experimentally in Ref.~\cite{DGG}. The setup consists of two metallic islands of finite capacitance $C$ (i.e., two floating Ohmic contacts) connected to each other by $N$ ballistic electronic channels. In addition, each island is coupled through $M$ ballistic channels to two macroscopic electrodes, as schematically shown in Fig.~\ref{Fig1}. The ballistic channels are implemented in a high-mobility Ga(Al)As two-dimensional electron gas tuned to the integer quantum Hall regime with filling factor $\nu$. The transparency of the contact between the floating islands and the quantum Hall edge channels is nearly perfect~\cite{Heat_CB}. The total number of incoming and outgoing edge channels is precisely controlled by quantum point contacts (QPCs), highlighted in yellow in Fig.~\ref{Fig1}. Accordingly, for the device studied in Ref.~\cite{DGG}, the number of available ballistic channels is constrained by $M\le 2\nu$ and $N\le \nu$.

The low-energy dynamics of the right- and left-moving ($\sigma$$=$$\pm$) quantum Hall edge states are described by chiral bosonic fields $\phi_{j,\sigma}(x,t)$, which obey the standard Kac-Moody commutation relation $[\partial_x\phi_{j,\sigma}(x,t),\phi_{j',\sigma'}(x',t)]=2i\sigma\pi\delta_{jj'}\delta_{\sigma\sigma'}\delta(x-x')$. The total Hamiltonian governing the edge dynamics and charging effects of both metallic islands consists of two contributions, $H=H_0+H_{\rm C}$. The first term,
\begin{eqnarray}\label{Ham}
H_{0}=\frac{\hbar v_F}{4\pi}\sum_{ j=1 }^{2M+N}\int_{-\infty}^{\infty} dx \left[\left(\partial_x \phi_{j,+}\right)^2 + \left(\partial_x \phi_{j,-}\right)^2\right],
\end{eqnarray}
represents the kinetic energy of the chiral edge channels, with $v_F$ the Fermi velocity. The second term accounts for the Coulomb blockade of the two metallic islands,
\begin{eqnarray}\label{CB}
H_{\rm C}=E_{\rm C}\frac{\left(\mathcal{Q}_1^2+\mathcal{Q}_{2}^2\right)}{e^2},
\end{eqnarray}
where $E_{\rm C}\equiv \pi \tau_{\rm C}^{-1}=e^2/2C$ denotes the charging energy~\cite{[{Since we consider a model without finite backscattering in the QPCs, we neglect the capacitive coupling between the two Ohmic contacts, described by the Hamiltonian term $\mathcal{Q}_1\mathcal{Q}_2/(e^2C_{12})$. Including this term would modify Eq.~(\ref{langevin}) (see Ref.~\cite{drag}) and, as a consequence, renormalize the gaps of the charged modes. For instance, one expects $\mathcal{M}_{\pm}E_{\rm C}\rightarrow (E_{\rm C}+C_{12}^{-1}/2)$ and $3(E_{\rm C}-C_{12}^{-1}/2)$ for the  particalur case $M=N=1$}]comment}. The operators $\mathcal{Q}_{1,2}$ describe the total charge accumulated on each island:
\begin{eqnarray}
\mathcal{Q}_{1(2)}= \sum_{\substack{j=1 \\ \sigma=\pm}}^{M(N)} \int_0^{\infty}dx \rho_{j,\sigma}(x)+\sum_{\substack{j=1 \\ \sigma=\pm}}^{N(M)} \int_{-\infty}^{0}dx \rho_{j,\sigma}(x),
\end{eqnarray}
where the charge-density operator is given by $\rho_{j,\pm}=\pm(e/2\pi)\partial_x\phi_{j,\pm}(x,t)$. To ensure proper regularization, we impose the boundary conditions
$\rho_{j,\pm}(\pm\infty)=0$, which reflect the decay of neutral excitations deep inside the Ohmic contacts.


{\it Theoretical approach.} Let us discuss a theoretical framework that allows us to calculate a heat flux carried by the outgoing edge states emitted from the metallic islands. We follow the Langevin equation approach, which we adopted from the seminal work of Slobodeniuk \textit{et al.}~\cite{Slobodeniuk}. 

Before dealing with the heat current, one needs to establish a correspondence between all incoming currents $j_{\rm in,\eta}^i$ and outgoing currents $j_{\rm out,\eta}^{i}$ (here, $i$ is an index of the edge state channel, $\eta=1,2$ denotes the left/right island, see~\cite{[{It is more convenient to work in the basis of the incoming/outgoing edge states rather than in the basis of left/right-moving ones. Here, $j^{i}_{\rm in,1}$ ($j^{i}_{\rm in,2}$) with $1\le i\le M$ (with $N<i\le M+N$) denotes the current incoming to the left (right) island from the left (right) reservoirs, while $j^{i }_{\rm in,1}$ ($j^{i }_{\rm in,2}$) with $M<i\le M+N$ (with $1\le i\le N$) is the incoming current from the right island}]footnote},~\cite{[{Details of the notations, geometry and derivations are shown in the Supplemental Material at [URL]}]suppl}). It can be done by solving the following system of equations:
\begin{subequations}
\begin{align}
    &\frac{d\mathcal{Q}_{\eta}(t)}{dt} = \sum_{i=1}^{M+N} \left\{ j_{\rm in,\eta}^i(t) - j_{\rm out,\eta}^i(t) \right\},\label{kirch}\\
    & j_{\rm out,\eta}^i(t) = \frac{\mathcal{Q}_{\eta}(t)}{R_qC} +  j_{\rm C,\eta}^i(t),\label{langevin}
\end{align}
\end{subequations}
where $R_q\equiv G_0^{-1}=2\pi\hbar/e^2$ is a quantum of resistance, and $j_{\rm C, \eta}^i$ is a Langevin current source. Equation~(\ref{kirch}) 
enforces charge conservation on the $\eta$-th metallic island. Equation~(\ref{langevin}) is a Langevin-type equation expressing that the $i$-th outgoing current is governed by two contributions: (i) the variation of the island's time-dependent electrostatic potential, $\mathcal{Q}_{\eta}(t)/C$, and (ii) the stochastic term 
$j_{\rm C,\eta}^i(t)$, which describes equilibrium current fluctuations of the island at the local temperature $T_{\rm C, \eta}$. The validity of the Langevin approach requires $T_{\textrm{C}, \eta}\gg \delta E$, where $\delta E$ denotes the energy level spacing of the Ohmic contacts, see details in Ref.~\cite{stabler2}.

Exploiting the correspondence between heat and charge current operators in a one-dimensional system with linear dispersion, one can present the heat flux in terms of the spectral density function of the current fluctuations $\delta j_{\rm out,\eta}^i = j_{\rm out,\eta}^i - \langle j_{\rm out,\eta}^i \rangle$, defined as
\begin{eqnarray}
2\pi\delta(\omega+\omega')S^{i}_{\textrm{out},\eta}(\omega)
= \langle \delta j^{i}_{\textrm{out},\eta}(\omega)
\delta j^{i}_{\textrm{out},\eta}(\omega') \rangle.
\end{eqnarray}
Therefore, the heat current carried by the $i$-th outgoing from $\eta$-th island edge channel yields the following form
\begin{eqnarray}\label{current_def}
J^{i}_{\rm out,\eta}=\frac{R_q}{4\pi }\int_{-\infty}^{\infty}d\omega
\left[S^{i}_{\rm out,\eta}(\omega)
-\frac{\hbar\omega \theta(\omega)}{R_{q}}\right],
\end{eqnarray}
where $\theta(\omega)$ is the Heaviside step function. The second term in Eq.~(\ref{current_def}) represents the subtraction of zero-point fluctuations present at zero temperature. 

Solving the system of Eqs.~(\ref{kirch}) and (\ref{langevin}) in the frequency domain allows us to express $S_{\rm out,\eta}^i(\omega)$ via the noise power of currents $j_{ p,\eta}^j(\omega)$ ($p=\{\textrm{in, C}\}$): 
\begin{eqnarray}\label{out}
S^{i}_{\textrm{out},\eta}(\omega)
= \sum_{p,\eta'} \sum_{j=1}^{M+N}
|\mathcal{T}^j_{p,\eta'}(\omega)|^2
S^{j}_{p,\eta'}(\omega).
\end{eqnarray}
Here, $\mathcal{T}_{p,\eta'}^j(\omega)$ can be interpreted as the scattering amplitude of bosonic density fluctuations~\cite{PhysRevB.105.075433}, and
the charge-current noise of the subsystem $p=\{\textrm{in, C}\}$, $2\pi\delta(\omega+\omega')S^{j}_{p,\eta}(\omega)=\langle \delta j^{j}_{p,\eta}(\omega)\delta j^{j}_{p,\eta}(\omega') \rangle $, exhibits equilibrium noise spectra at their respective local temperature $T_{p, \eta}$:
\begin{eqnarray}\label{equilibrium}
S^j_{p,\eta}(\omega)
= \frac{\hbar\omega G_0}{1-e^{-\hbar\omega/k_BT_{p,\eta}}}.
\end{eqnarray}


We next evaluate the heat current, Eq.~(\ref{current_def}), using Eqs.~(\ref{out}) and (\ref{equilibrium}) for two distinct experimental configurations that can be realized within the same double-island device: (i) the heat Coulomb blockade and (ii) the thermal transport configuration, corresponding to a temperature-biased setup between source and drain electrodes.

\textit{Heat Coulomb blockade.} We first consider the configuration of the pioneering heat Coulomb blockade experiment~\cite{Heat_CB}. Let us assume that all electrodes have the same base temperature, $T_{\rm in,1}$$=$$T_{\rm in,2}$$=$$T_{\rm in}$. Meanwhile, a dc bias voltage is applied symmetrically to both left and right source electrodes. The pumped energy dissipates through Joule heat, resulting in metallic islands heating up to a steady electronic temperature $T_{1}$$=$$T_{2}$$=$$T_{\rm C}$$>$$T_{\rm in}$. Under these conditions, Eq.~(\ref{out}) simplifies considerably once the corresponding temperatures are assigned to the different channels~\cite{suppl}.

Summing heat flow Eq.~(\ref{current_def}) in all outgoing edge channels, $J_Q=\sum_{i=1}^{M}\left[J^{i}_{\textrm {out},1}+J^{N+i}_{\textrm {out},2}-(\pi /6\hbar)k_B^2T_{\rm in}^2\right]$,
we obtain the total heat current:
\begin{eqnarray}\label{HCB}
&&J_{Q}=2M\left\{\frac{\pi k_B^2}{12\hbar}(T_{\rm C}^2-T_{\rm in}^2)\right.\nonumber\\
&&\left.+\frac{ME_{\rm C}^2}{4\pi^3\hbar}\sum_{\alpha=\pm}\left[\mathcal{F}\left(\frac{\mathcal{M}_{\alpha}E_{\rm C}}{\pi k_BT_{\rm in}}\right) -\mathcal{F}\left(\frac{\mathcal{M}_{\alpha}E_{\rm C}}{\pi k_BT_{\rm C}}\right)\right]\right\},
\end{eqnarray}
where $\mathcal{F}(x)$ is a dimensionless integral function that has the following form:
\begin{eqnarray}\label{f_function}
&&\mathcal{F}(x)\equiv\int_0^{\infty}dz
\frac{z}{z^2+x^2}\frac{1}{e^{z}-1}\nonumber\\&&~~~~~~~~=\frac{1}{2}\left[\log\left(\frac{x}{2\pi}\right)-\frac{\pi}{x}-\psi\left(\frac{x}{2\pi}\right)\right],
\end{eqnarray}
and $\psi(x)$ is a digamma function. In Eq.~(\ref{HCB}), the parameters $\mathcal{M}_-=M$ and $\mathcal{M}_+=2N+M$
serve as effective "masses" of two gapped collective modes associated with charge dynamics on the islands. The model defined by Eqs.~(\ref{Ham}) and (\ref{CB}) can be diagonalized into normal modes (see, e.g., Refs.~\cite{parafilo_2023,parafilo_fnt}), yielding $2(2M+N)-2$ gapless neutral modes corresponding to charge-conserving excitations of the edge channels, and two gapped (charged) modes with energy gaps proportional to $ \mathcal{M}_{\pm}E_{\rm C}/\pi^2$. 


Using the asymptotic of integral function Eq.~(\ref{f_function}), $\mathcal{F}(x\gg1)\approx \pi^2/6x$, we obtain the total heat current in the limit $T_{\rm C}, T_{\rm in}\ll \mathcal{M}_{\alpha}E_{\rm C}/k_B$: 
\begin{eqnarray}\label{HCB1}
J_{Q}=\frac{\pi k_B^2}{12\hbar}\left(T_{\rm C}^2-T_{\rm in}^2\right)\left[2M-1-\frac{M^2}{(2N+M)^2}\right].
\end{eqnarray}
Equation~(\ref{HCB1}) represents the first main result of this Letter and shows a clear difference from the conventional heat Coulomb blockade~\cite{Slobodeniuk,Heat_CB}. Unlike the single-island configuration, where the heat flux is reduced by precisely one quantum of thermal conductance, here we find a more subtle suppression factor explicitly dependent on $N, M$. 

The result in Eq.~(\ref{HCB1}) can be understood by considering two limiting cases. In the regime $N\gg M$, the two islands effectively merge into a single island, and Eq.~(\ref{HCB1}) reduces to the conventional heat Coulomb blockade, where one channel out of $2M$ is suppressed. Conversely, for $M\gg N$, the inter-island coupling becomes negligible, and the system behaves as two independent islands, each exhibiting its own heat Coulomb blockade.

Note that the number of inter-island channels 
$N$ appears only in the suppression factor. This is a consequence of the device geometry: the structure consists of two coupled subparts, and the
$N$ channels do not directly contribute to the measured heat current. However, they affect the result indirectly, as becomes evident in the diagonalized eigenmode basis.
Coulomb interactions Eq.~(\ref{CB}) mix the edge channels, giving rise to collective charge modes delocalized over both islands. As a result, the eigenmodes responsible for heat transport between each island and its drain contain admixtures of the inter-island channels, leading to an effective suppression of the total heat current that can be expressed through the reduced mass parameter $M^2\left(\mathcal{M}_{+}^{2}+\mathcal{M}_{-}^{2}\right)/\mathcal{M}_{+}^{2}\mathcal{M}_{-}^{2}$.

\begin{figure}
\centering 
\includegraphics[width=1\columnwidth]{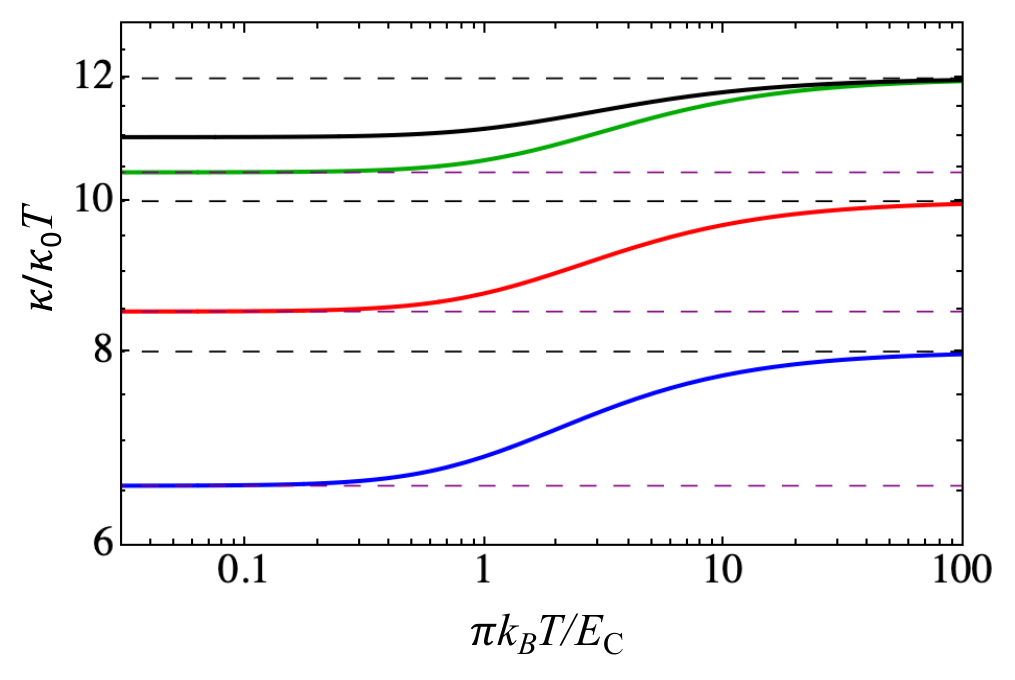} \caption {Normalized heat conductance $\kappa/\kappa_0T_{\rm C}$ from Eq.~(\ref{hconductance}) as a function of dimensionless temperature $\tau_{\rm C} k_BT_{\rm C}=\pi k_BT_{\rm C}/E_{\rm C}$. Blue, red, and green solid lines indicate the heat conductance for $N=1$ and $M=4,5,6$, respectively. The black line indicates the case of $M=6$ and large $N\gg M$. Black and purple dashed lines depict asymptotic limits of $2M$ and $2M$$-$$1$$-$$M^2/(M+2N)^2$, respectively.}
\label{Fig2} 
\end{figure}

To demonstrate the lifting of the heat Coulomb blockade with increasing temperature, it is more convenient to consider the thermal conductance rather than the heat current. Within the linear response, we take $T_{\rm C}=T+\Delta T$, $T_{\rm in}=T$, and expand Eq.~(\ref{HCB}) to the leading order in $\Delta T$. As a result, the heat conductance, defined as $\kappa = \partial J_Q/\partial \Delta T$, takes the form [here we set $\hbar$$=$$k_B$$=$$1$ to simplify Eq.~(\ref{hconductance})]
\begin{eqnarray}\label{hconductance}
&& \kappa = 2M\kappa_0 T   + \frac{M^2}{2\pi \tau^2_{\rm C} T}  \nonumber  \\
&&+M^2\sum_{\alpha=\pm}\left[\frac{1}{4 \tau_{\rm C} \mathcal{M}_{\alpha}}\right.\left.-\frac{\mathcal{M}_{\alpha}}{8\pi^2 \tau_{\rm C}^3T^2}\psi'\left(\frac{\mathcal{M}_{\alpha}}{2\pi \tau_{\rm C} T}\right)\right],
\end{eqnarray}
where $\kappa_0=\pi^2 k^2_B/3h$, and, thus, $\kappa_0 T$ is the quantum of thermal conductance. The temperature dependence of heat conductance Eq.~(\ref{hconductance}) for different numbers of ballistic channels is shown in Fig.~\ref{Fig2}. The crossover between $2M-1-(\mathcal{M}_-/\mathcal{M}_+)^2$ and $2M$ asymptotics is clearly seen with temperature change between $T\ll E_{\rm C}/k_B$ and $T\gg E_{\rm C}/k_B$ regimes.

\textit{Transport regime.} We now consider a 
typical configuration of the heat transport experiment in which both metallic islands are positioned between the heat source and drain reservoirs. Accordingly, we set $T_{\rm in,1}=T_S$ and $T_{\rm in, 2}=T_D$ ($T_S$$>$$T_D$), see Fig.~\ref{Fig1}. The quasi-equilibrium temperatures of the islands $T_1$, $T_2$ are not externally fixed and should be obtained self-consistently using the relations between incoming and outgoing heat fluxes. 

To further simplify the problem, we study only the regime of heat Coulomb blockade, $T_{S(D)}, T_{1(2)}\ll E_{\rm C}/k_B$. Using Eqs.~(\ref{current_def}),~(\ref{out}), and (\ref{equilibrium}), we determine a total heat current through the system in terms of the net heat flow between the two islands, $J_Q=\sum_{i=1}^{N} \left[J^{M+i}_{\rm out,1}-J^i_{\rm out,2}\right]$.
After straightforward calculations~\cite{suppl}, one gets the heat current
\begin{eqnarray}\label{heat_transport}
J_Q=\frac{N\kappa_{0}}{2}\left\{\frac{T_S^2-T_D^2}{2N+M}+\frac{2N-1+M}{2N+M}\left(T_1^2-T_{2}^2\right)\right\},
\end{eqnarray}
which depends not only on the imposed temperature bias $T_S-T_D$, but also on the local temperatures of the two intermediate Ohmic contacts.

Next, we write the energy balance equations expressing conservation of the total incoming and outgoing heat flux for each metallic island:
\begin{eqnarray}\label{system}
\begin{array}{c}
   \sum_{i=1}^{N} J^{M+i}_{\rm in,1}+\frac{M\kappa_0}{2}T_S^2=\sum_{j=1}^{M+N}J^j_{\rm out,1},  \\
    \sum_{i=1}^{N} J^{i}_{\rm in,2}\,\,\,+\frac{M\kappa_0}{2}T_D^2=\sum_{j=1}^{M+N}J^j_{\rm out,2}.
\end{array}
\end{eqnarray}
Local island temperatures $T_1$, $T_2$ are determined by solving Eq.~(\ref{system}) within the heat Coulomb blockade regime~\cite{suppl}, which yields the following result:
\begin{eqnarray}\label{temperature}
T_{1(2)}^2=T_{\rm m}^2\pm\frac{(2N+M)(M-1)\left(T_S^2-T_D^2\right)}{2\left[M^2+M(4N-1)+4N(N-1)\right]},
\end{eqnarray}
where $T_{\rm m}^2=(T_S^2+T_D^2)/2$ denotes the mean temperature of the device. In the limit of strong inter-island coupling, $N \gg M$, the two metallic islands effectively merge into a single one, and the local temperatures $T_{1,2}$ approach the mean temperature of the system. Contrary, when $M\gg N$, each island is well thermalized with its adjacent reservoir, having $T_{1(2)}\rightarrow T_{S(D)}$~\cite{[{Note that a similar "thermalization" is expected to occur in an asymmetric single-island setup when it is coupled to the heat source via one channel and to the heat drain via 
$M\equiv N-1$ channels. In this case, the local island's temperature is $T_{\rm C}^2=[2 T_S^2+(N^2-2)T_D^2]/N^2$, while the heat current reads as $J_Q=(\kappa_0/2)(T_S^2-T_D^2)M^2(M+3)/(M+1)^3$, see~\cite{suppl}}]single}. 
At $M=N$, the local quasi-equilibrium temperatures $T_{1,2}$ are given by
\begin{eqnarray}\label{temperature2}
    T^2_{1(2)}= \frac{6M-4}{9M-5}T_{S(D)}^2+\frac{3M-1}{9M-5}T_{D(S)}^2.
\end{eqnarray}
At a large number of channels, $M\rightarrow \infty$, Eq.~(\ref{temperature2}) reduces to the temperature profile of the non-interacting system, $E_C=0$. 

In the special case $M=1$ and arbitrary $N$, the local island temperatures $T_{1,2}$ become exactly equal to the mean temperature $T_{\rm m}$. This counterintuitive behavior, predicted in Refs.~\cite{roche,sela}, reflects a subtle interplay between charge and neutral excitations in the multiple-island hybrid metal-semiconductor device. 
As discussed in Ref.~\cite{roche}, for $M=1$ the islands thermally decouple from the reservoirs because neutral modes cannot be emitted by the source or drain in the presence of only a single outgoing channel, whereas charge modes are suppressed by the heat Coulomb blockade.



Equations~(\ref{heat_transport}) and (\ref{temperature}) together constitute the second main result of this work, establishing the complete temperature and heat transport characteristics of the double-island system. These results provide a framework for testing the validity of the Wiedemann-Franz law in the hybrid metal-semiconductor device.

\textit{Lorenz ratio.} The Wiedemann-Franz (WF) law expresses a fundamental relation between electronic heat and charge transport at low temperatures. In mesoscopic systems, it states that the ratio of the thermal conductance $\kappa$ to the product of electrical conductance $G$ and temperature $T$ equals a universal constant $\mathcal{L}_0=(\pi^2/3)(k_B/e)^2$ -- the Lorenz number. Deviations from this universal relation, quantified by the Lorenz ratio $\mathcal{L}=\kappa/(G T)$, are commonly regarded as a hallmark of NFL behavior or of strong interactions that decouple charge and heat transport~\cite{PhysRevLett.100.066801,kanefisher,krive,krive2,kiselev_WF}. 

We characterize the violation of the WF law by the dimensionless parameter $\mathcal{R}=\lim_{T\rightarrow 0} \mathcal{L}/\mathcal{L}_0$. Let us first consider the special limit of arbitrary $N$ and $M=1$ when $T_1$$=$$T_2$. Linearizing Eq.~(\ref{heat_transport}) with respect to temperature ($T_S=T+\Delta T$, $T_D=T$) and using the maximal charge conductance at the NFL critical point from Ref.~\cite{PhysRevB.110.L241406} (see Eq.~(13) therein), we obtain
\begin{eqnarray}\label{ins}
 \kappa=\frac{N}{2N+1}\frac{\pi^2k_B^2 T}{3h}\quad,\quad G = \frac{N}{2N+1} \frac{e^2}{h},
\end{eqnarray}
and, thus, $\mathcal{R}=1$. Therefore, the double-island hybrid device, operating in the regime of the Luttinger liquid simulator~\cite{PhysRevB.110.L241406}, satisfies the WF law.
In the limit $N\rightarrow \infty$, where the two Coulomb islands effectively merge into a single island, Eq.~(\ref{ins}) coincides with the result for a two-channel charge Kondo simulator in the low-temperature regime $T\ll T_K$ (where $T_K$$\propto $$E_{\rm C}$ is the Kondo temperature of the charge Kondo device): $\kappa=\kappa_0 T/2$ and $G=G_0/2$, as obtained in Ref.~\cite{fritz} using the Emery-Kivelson model.
In the limit, $N=1$, one recovers the heat and charge conductance characteristics of the 2SCK circuit: $\kappa=\kappa_0T/3$ (see Ref.~\cite{sela}) and $G=G_0/3$~\cite{two-site}.

\begin{figure}[t!]
\centering 
\includegraphics[width=1\columnwidth]{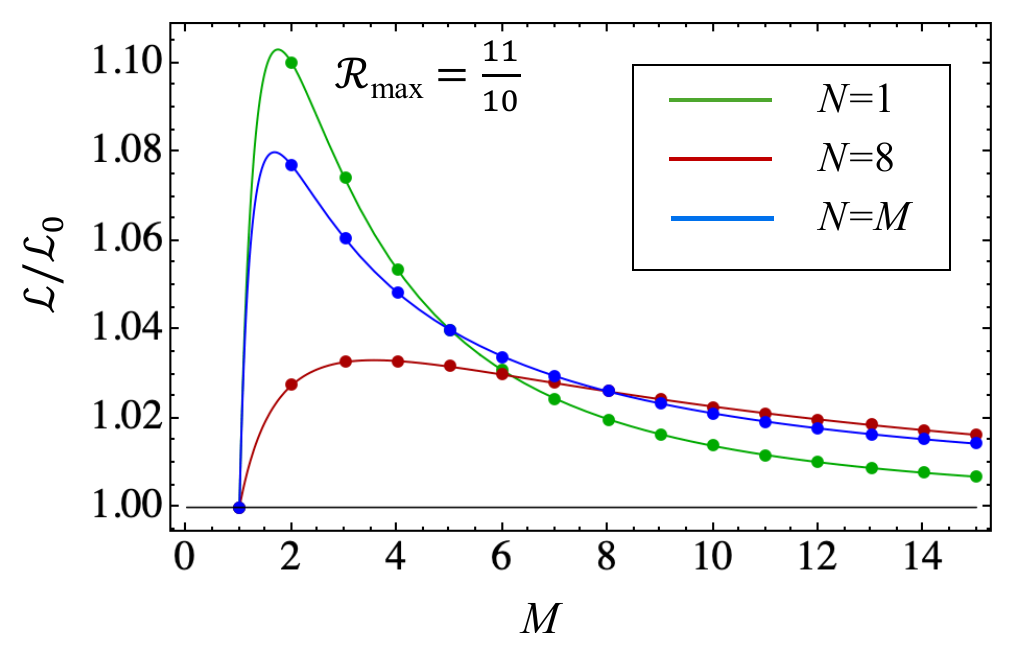} \caption {Normalized Lorenz ratio $\mathcal{R}=\mathcal{L}/\mathcal{L}_0$ from Eq.~(\ref{lorenz_ratio}) as a function of the number of ballistic channels $M$. The maximal possible value of Lorenz number $\mathcal{R}_{\rm max}=11/10$ occurs for $N=1$, $M=2$.}
\label{Fig3} 
\end{figure}

Ohmic contact-based hybrid setups, while serving as building blocks of charge Kondo simulators~\cite{pierre_2CCK,pierre_3CCK,DGG}, cannot be fully described as the charge Kondo model unless quantized charge states are established on the island(s) -- a condition ensured by finite backscattering in the QPCs. Nevertheless, the conductance behavior in the above examples is reminiscent of that expected at the NFL fixed points of some charge Kondo models in the low-temperature limit $T/T_K\rightarrow 0$. This correspondence indicates that the conclusions of Refs.~\cite{fritz,PhysRevB.102.241402}, demonstrating the validity of the WF law at the NFL fixed points of two-channel Kondo models, can be "extended", at least to some extent, to certain realization of multi-channel single- and two-site charge Kondo simulators~\cite{[{In addition to the above cases of $M=1$ and arbitrary $N$, $N\rightarrow \infty$, $N=1$, one may speculate on the behavior of the WF law in a $2M$-channel charge Kondo circuit in the vicinity of the NFL critical point at $T/T_K\rightarrow 0$. Using Eqs.~(\ref{heat_transport}) and (\ref{temperature}) in the limit $N\rightarrow \infty$ at arbitrary $M$, we obtain $\kappa=(M/2)\kappa_0T$ and $G=(M/2)G_0$, which again yields $\mathcal{R}=1$}]extra}.

Returning to the more general case, we now evaluate the Lorenz ratio for any $M$ and $N$ by using Eqs.~(\ref{heat_transport}), (\ref{temperature}) together with the electrical conductance $G=G_0MN/(2N+M)$ calculated according to the scheme described, e.g., in~\cite{parafilo,parafilo_2023}. After straightforward algebra, we arrive at
\begin{eqnarray}\label{lorenz_ratio}
   \mathcal{R}=\frac{1}{M}\left\{1+\frac{(2N - 1 + M)(M + 2 N)(M-1)}{M^2 + 4 N (N-1) + M (4N -1 )}\right\}.
\end{eqnarray}
Figure~\ref{Fig3} illustrates the Lorenz ratio obtained from Eq.~(\ref{lorenz_ratio}) as a function of $M$ for several fixed $N$. The ratio exhibits a pronounced nonmonotonic dependence, bounded between $\mathcal{R}_{\rm max}=11/10$ and $\mathcal{R}_{\rm min}=1$.
The upper bound occurs for $M=2$, $N=1$, indicating the strongest deviation from the WF law, whereas the lower bound corresponds to $M=1$ or to the limit $M\rightarrow\infty$ (or $N\rightarrow\infty$), where conventional Fermi-liquid behavior is restored. The channel dependence predicted by Eq.~(\ref{lorenz_ratio}) is consistent with the Lorenz ratio found for the chain of small metallic islands (see Ref.~\cite{roche} and Fig.~4 (b) therein). The similar behavior also appears in an asymmetric single-island device coupled to the heat source and drain via one and $M\equiv N-1$ channels, respectively, yielding $\mathcal{R}=M(3+M)/(1+M)^2$~\cite{suppl}.


We also urge the reader to compare Eq.~(\ref{lorenz_ratio}) with the result of Ref.~\cite{kiselev_WF}, where the Lorenz ratio was obtained for the same double-island hybrid metal-semiconductor architecture but with a tunnel barrier separating two metallic islands. 

\textit{Conclusions.} We studied the heat Coulomb blockade in a system with two gapped (charged) modes by considering a double-island hybrid metal-semiconductor device.
Our results reveal that the suppression of heat flux in this configuration is more subtle than in the conventional single-island case: The magnitude of suppression is determined by the number of ballistic channels connecting the islands to each other and to the reservoirs. In addition, we examine an alternative configuration in which the double islands are positioned between a heat source and a drain. We showed that the imbalance between the inter-island channels and those linking the islands to external reservoirs dictates both the thermal current and the degree of equilibration between the islands. By evaluating the Lorenz ratio, we demonstrated a controllable violation of the Wiedemann-Franz law, thereby providing access to nontrivial regimes of mesoscopic heat transport.

\textit{Acknowledgements.} The author would like to especially thank B. Altshuler, D. X. Nguyen, and A. Andreanov for fruitful discussions. The author also acknowledges support from the Institute for Basic Science in Korea (IBS-R024-D1).

\vspace*{5mm}

\bibliography{biblio}


\end{document}